\documentclass[twocolumn,pra,aps,superscriptaddress,showpacs,eqsecnum]{revtex4-1}
\makeatletter
\def\@bibdataout@aps{%
 \immediate\write\@bibdataout{%
  @CONTROL{%
   apsrev41Control,author="08",editor="1",pages="0",title="0",year="1",eprint="1"%
  }%
 }%
 \if@filesw
  \immediate\write\@auxout{\string\citation{apsrev41Control}}%
 \fi
}%
\makeatother 

\usepackage{yfonts} 
\usepackage[usenames,dvipsnames]{xcolor}
\usepackage{amsmath,amsfonts,mathrsfs,amssymb}
\usepackage{tikz}
\usepackage{graphicx}
\usepackage{epstopdf}
\usepackage{epsfig}
\usepackage{color}
\usepackage{soul} 
\usepackage{wasysym}
\usepackage[breaklinks=true]{hyperref}

\hypersetup{
  colorlinks   = true, 
  urlcolor     = blue, 
  linkcolor    = blue, 
  citecolor   = red 
}
\usepackage{url}

\newcommand{\smallfrac}[2]{\mbox{$\frac{#1}{#2}$}}
\newcommand{\half}{\smallfrac{1}{2}}

\newcommand{\bra}[1]{\left\langle{#1}\right|}
\newcommand{\ket}[1]{\left|{#1}\right\rangle}

\newcommand{\expt}[1]{\!\left\langle{#1}\right\rangle}

\newcommand{\sE}{\mathcal{E}}

\newcommand{\tr}{\mbox{tr}}

\newcommand{\mytilde}{\raise.17ex\hbox{$\scriptstyle\mathtt{\sim}$}}

\newcommand{\dg}{^\dag}

\def\Pcheck{p_{\checkmark}}

\usepackage{cleveref}

\begin{document}
\title{Models of reduced-noise, probabilistic linear amplifiers}

\author{Joshua Combes}
\affiliation{Institute for Quantum Computing, Department of Applied Mathematics, University of Waterloo, Waterloo, ON, Canada}
\affiliation{Perimeter Institute for Theoretical Physics, 31 Caroline St.~N, Waterloo, Ontario, Canada N2L 2Y5}
\affiliation{Centre for Engineered Quantum Systems, School of
Mathematics and Physics, University of Queensland, Brisbane,
Queensland 4072, Australia}

\author{Nathan Walk}
\affiliation{Centre for Quantum Dynamics, Griffith University, Nathan, Queensland 4111, Australia}
\affiliation{Centre for Quantum Computation and Communication Technology, School of
Mathematics and Physics, University of Queensland, Brisbane,
Queensland 4072, Australia}
\affiliation{Department of Computer Science, University of Oxford, Wolfson Building, Parks Road, Oxford OX1 3QD, United Kingdom}

\author{A.~P.~Lund}
\affiliation{Centre for Quantum Computation and Communication Technology, School of
Mathematics and Physics, University of Queensland, Brisbane,
Queensland 4072, Australia}
\author{T.~C.~Ralph}
\affiliation{Centre for Quantum Computation and Communication Technology, School of
Mathematics and Physics, University of Queensland, Brisbane,
Queensland 4072, Australia}
\author{Carlton M.~Caves}
\affiliation{Center for Quantum Information
and Control, University of New Mexico, Albuquerque, NM 87131-0001,
USA}
\affiliation{Centre for Engineered Quantum Systems, School of
Mathematics and Physics, University of Queensland, Brisbane,
Queensland 4072, Australia}

\begin{abstract}
We construct an amplifier that interpolates between a nondeterministic, immaculate linear amplifier and a deterministic, ideal linear amplifier and beyond to nonideal linear amplifiers. The construction involves cascading an immaculate linear amplifier that has amplitude gain $g_1$ with a (possibly) nonideal linear amplifier that has gain $g_2$.  With respect to normally ordered moments, the device has output noise $\mu^2(G^2-1)$ where $G=g_1 g_2$ is the overall amplitude gain and $\mu^2$ is a noise parameter.  When $\mu^2\ge1$, our devices realize ideal ($\mu^2=1$) and nonideal ($\mu^2>1$) linear amplifiers.  When $0\le\mu^2<1$, these devices work effectively only over a restricted region of phase space and with some subunity success probability $\Pcheck$.  We investigate the performance of our $\mu^2$-amplifiers in terms of a gain-corrected probability-fidelity product and the ratio of input to output signal-to-noise ratios corrected for success probability.
\end{abstract}

\date{\today}

\maketitle

\section{Introduction and motivation}
\label{sec:intro}

Quantum-limited amplification is an important method for probing the
microscopic world. The canonical quantum amplifier is called a phase-preserving
linear quantum amplifier.  It takes an input bosonic signal and
produces a larger output signal~\cite{Haus1962a,Caves1982a}, while preserving
the phase. The quantum constraints on the operation of such a device are
ultimately a consequence of unitarity and can be thought as coming from the
prohibition on transformations that increase the distinguishability of
nonorthogonal states~\cite{WooZur82,Dieks82}. Until recently the only
constraint known was a bound on the second moment of added noise, which is
tight if the added noise is Gaussian~\cite{Haus1962a,Caves1982a}. In
Ref.~\cite{CavComJiaPan12}  in-principle constraints for all moments of added
noise were worked out in detail. An amplifier that adds the  minimum amount of
noise allowed by quantum theory is called an {\em ideal\/} linear
amplifier; one that adds more than the minimum is called a {\em nonideal\/}
linear amplifier.

Ralph and Lund~\cite{RalLun09} and, independently, Fiur\'a\v{s}ek~\cite{Fiu04}
(in the context of cloning) proposed an intriguing idea that suggested it
might be possible to build an amplifier that {\em subtracts noise\/}! This
proposal continues to generate interest from the
community~\cite{XianRalpLund10,EleBarJef13,TonRad13,WalkLundRalp13,ChrWalAss14,BlaBarGra15,DonColEle15,ChirYangHuan15,ParkJooZava16,RosaMariGiov16}, with potential applications including quantum key distribution \cite{Fiurasek:2012p7670,Walk:2013p403,Blandino:2012p5681} and the distillation of quantum correlations \cite{ChrWalAss14}. Specifically, the
proposed amplifier with amplitude gain $g>1$ takes an input coherent state
$|\alpha\rangle$ to a ``target'' coherent state $|g\alpha\rangle$ with (success)
probability $\Pcheck$ and fails with probability $1-\Pcheck$.
This device amplifies the normally ordered input noise, so when applied to an input coherent state
to produce an output coherent state, neither of which has any normally ordered noise, the device adds no 
noise as measured by normally ordered moments.  Hence it was originally called a \emph{nondeterministic noiseless\/} linear amplifier or NLA~\cite{RalLun09}.  When compared to a classical noiseless amplifier, however, such a device is actually
better than noiseless. A classical noiseless amplifier would amplify the symmetrically ordered input noise to the output without the addition of any noise; this device has been called a {\em perfect amplifier\/}~\cite{ComCav13}.
Because the device proposed by Ralph and Lund and by Fiur\'a\v{s}ek is better than
perfect, it has been christened an {\em immaculate amplifier} in Ref.~\cite{ComCav13}.

There is, in fact, a continuum of devices between an ideal linear amplifier
and an immaculate amplifier; all the amplification devices
in this continuous family, except the ideal linear amplifier, work probabilistically.
The objective of this article is to explore the properties of this family of devices.  Any
amplification device that works probabilistically can be called, following Ralph and Lund's original
terminology~\cite{RalLun09}, a nondeterministic
linear amplifier or NLA; the family of devices we study in this paper is a particular subset
of such nondeterministic linear amplifiers.  It is worth noting that although our analysis
of such devices focuses on coherent-state inputs, the devices can be applied to any input state.

The quantum limits for such devices are not usually characterized by the
amount of added noise.  Instead, they are characterized by three properties:
the operating region of phase space over which the device can amplify input
coherent states effectively, the success probability $\Pcheck$,  and the
fidelity to the ``target'' coherent state.  If the input region is taken to
be the entire phase plane and the fidelity to the target state is one---i.e.,
an immaculate amplifier that works on the entire phase plane---the probability
that such a device works is strictly zero~\cite{MenCro09}.  Even should one
restrict the input coherent states to a circle in phase space centered at
the origin, if one demands unit fidelity to the amplified target state, one
can show that the success probability is zero~\cite{ComCav13}.

If one sticks with phase-preserving amplification, one must both restrict
the phase-space region over which the amplifier is supposed to work {\em and\/}
loosen the fidelity-one requirement. For example, for the optimal model of an
immaculate linear amplifier, when the input coherent states are restricted to
the disk of complex amplitudes $|\alpha|<\sqrt{N}/g$, where $N$ is a
number-basis cutoff, the fidelity of the amplifier output to $|g\alpha\rangle$
is $F\simeq 1$, but the success probability scales as
$\Pcheck \simeq e^{-|\alpha|^2}/g^{2N}$.  These quantum limits
are known to be tight~\cite{ComCav13} and physically realizable~\cite{ComCav13,McMLunRal14}.
All of the analysis in this paper is from the perspective of how well these probabilistic 
devices work when the effect of failed attempts is included, a perspective that is appropriate 
for many metrology and communication tasks. In other situations, it is often advantageous 
only to only keep successful amplification attempts \cite{CombFerr15}.

The purpose of this paper is to give a constructive method for building physical
devices that interpolate between the ideal linear amplifier and the immaculate
amplifier and, since it is easy to include in the formalism, between the ideal
linear amplifier and noisier, nonideal linear amplifiers.  For the case of the perfect amplifier,
investigations along these lines were originally suggested by a subset of the current authors \cite{RalLunWal14}.
In Sec.~\ref{sec:laprior}, we review the description of physical and unphysical amplifiers, ranging from
unphysical immaculate and perfect amplifiers to the physical ideal amplifier and
then beyond to physical nonideal amplifiers.  In giving this description, a natural
parameter $\mu^2$ arises to characterize the amount of noise added to or subtracted
from the output.  The boundary between unphysical and physical amplifiers is at the
ideal linear amplifier, which corresponds to $\mu^2=1$.  The unphysical amplifiers, which
add less noise than the ideal amplifier, correspond to $0\le\mu^2<1$, and the physical,
nonideal amplifiers that add more noise than the ideal linear amplifier correspond to
$\mu^2>1$ .  In Sec.~\ref{sec:laprior} we also review an uncertainty-principle bound
that restricts the success probability of the unphysical, $\mu^2<1$ amplifiers when
they are made physical.

\begin{figure}[t]
\includegraphics[width=\columnwidth]{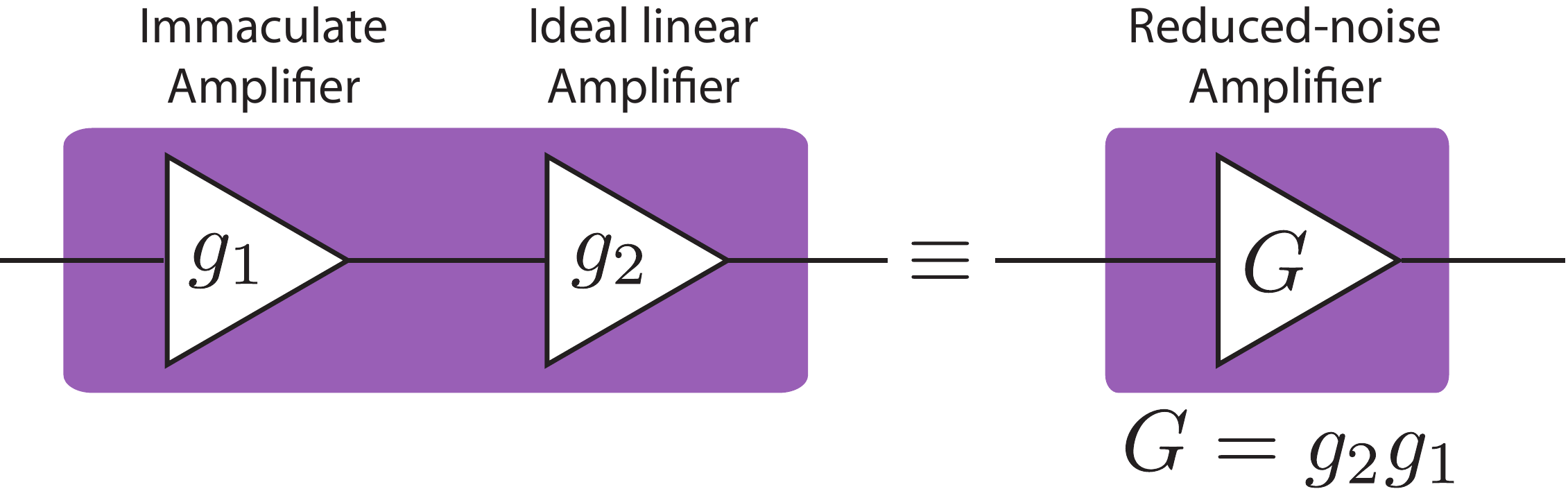}
\caption{(Color online).  A  family of reduced-noise amplifiers, which we call $\mu^2$-amplifiers, can be constructed
by cascading an optimal physical model of an immaculate amplifier (amplitude gain $g_1$) with a physical linear
amplifier (amplitude gain $g_2$) to achieve a reduced-noise amplifier with gain $G=g_2g_1$.}\label{fig1}
\end{figure}

In Sec.~{\ref{subsec:immac} we show how to construct a physical family of $\mu^2$-amplifiers
by cascading an optimal physical model of an immaculate amplifier with a physical amplifier (see Fig.~\ref{fig1});
the best $\mu^2$-amplifiers result from making the physical amplifier in this construction ideal.  In Sec.~\ref{subsec:upbounds} we investigate the performance of our physical $\mu^2$-amplifiers when they are operating in the high-fidelity operating region; performance is measured in terms of a probability-fidelity product and the noise figure.  The approximations made for operation within the high-fidelity operating region are dropped in Sec.~\ref{sec:exact_n_equals_1},
where we explore exact measures of performance for operation both within and outside the
high-fidelity operating region. Section~\ref{sec:con} provides concluding thoughts.

\section{Physical and unphysical linear amplifiers} \label{sec:laprior}
We assume the reader is familiar with the mathematics of linear amplification, immaculate amplification, and quasiprobability
distributions. Pedagogical material is available in Refs.~\cite{Caves1982a,ComCav13,CavComJiaPan12}.

The setting for our investigation is a signal carried by a single-mode field,
\begin{align}
E(t)&
\!=\!\frac{1}{2}(ae^{-i\omega t}+a^\dagger e^{-i\omega t})
\!=\!\frac{1}{\sqrt2}(x_1\cos\omega t+x_2\sin\omega t)\,.
\end{align}
This {\em primary mode}, which we label by $A$, is to undergo phase-preserving
linear amplification.  The annihilation and creation operators, $a$ and
$a^\dagger$, are related to the Hermitian quadrature components, $x_1$ and
$x_2$, by
\begin{align}\label{eq:x1x2}
a=\frac{1}{\sqrt2}(x_1+ix_2)\,,\quad a^\dagger=\frac{1}{\sqrt2}(x_1-ix_2)\,,
\end{align}
where $[a,a^\dagger]=1$ or, equivalently, $[x_1,x_2]=i$.

In Ref.~\cite{CavComJiaPan12} is was shown that the action of any phase-preserving
linear amplifier on an input state $\rho$ of the primary mode can be represented by
a map $\sE$ such that
\begin{align}\label{eq:E}
\rho_{\rm out}=\sE(\rho)={\rm Tr}_{B}[S(r)\rho\otimes\sigma S^\dagger(r)]\,.
\end{align}
In this expression, $\sigma$ is the input state of a (perhaps fictitious)
ancillary mode $B$, which has annihilation and creation operators $b$ and
$b^\dagger$, and $S(r)=e^{r(ab-a^\dagger b^\dagger)}$ is the two-mode squeeze
operator.  The amplitude gain is given by $G=\cosh r$, and the noise
properties of the amplifier are encoded in the ancillary state~$\sigma$.

Inspired by this general description of linear amplifiers, Ref.~\cite{ComCav13}
pointed out that all linear amplifiers from immaculate to nonideal can be
characterized by a sequence of maps for which the ancilla states are Gaussian
states of thermal form,
\begin{align}
\sigma(\mu^2)
=\frac{1}{\mu^2}\!\left( 1 -\frac{1}{\mu^2} \right)^{a\dg a}
=\frac{1}{\mu^2}\sum_{n=0}^\infty\left(1-\frac{1}{\mu^2}\right)^n \ket{n}\bra{n} \,.
\label{eq:sigmamu}
\end{align}
When $\mu^2\in[1,\infty)$, $\sigma$ is a physical thermal state, with
dimensionless inverse temperature $\beta$ given by
$\mu^2=(1-e^{-\beta})^{-1}=\bar n+1$, where $\bar n = \tr[ b\dg b\sigma]$
is the mean number of quanta; $\mu^2=1$ gives the vacuum state.  When
$\mu^2\in[0,1)$, $\sigma$ has negative eigenvalues and thus is
unphysical (notice that $\bar n=\mu^2-1<0$).  The amplifier maps corresponding to
these unphysical $\sigma$  are not completely positive and thus are also unphysical~\cite{CavComJiaPan12}.

We now focus attention on four types of phase-preserving amplifiers, which correspond
to various values of $\mu^2$:
\begin{enumerate}
\item[0.]
The {\em nonideal linear amplifier} (physical), which corresponds to $\mu^2>1$.
\item[1.]
The {\em ideal linear amplifier} (physical), which corresponds to $\mu^2=1$.
\item[2.]
The {\em perfect linear amplifier} (unphysical), which corresponds to $\mu^2=1/2$.
\item[3.]
The {\em immaculate linear amplifier} (unphysical), which corresponds to $\mu^2=0$.
\end{enumerate}
The function of these four amplifiers
can be understood intuitively in terms of how the output noise arises from
amplified input noise and added noise. Cahill and Glauber's $s$-ordered
quasiprobability distributions~\cite{Cahill1969a} provide a natural way of
understanding the relationship between input and output noise.

In this paper we consider the action of amplifiers on coherent  states.
Of course, the amplifier maps can be applied to any input state, but
coherent states, due to their minimal phase-insensitive Gaussian noise, are useful
for elucidating the properties of amplifier maps. In particular, we compare
the first and second moments of the input and output of the amplifiers.

For input coherent state $\ket{\alpha}$, the mean complex amplitude and
$s$-ordered variance of the input are
\begin{align}
\expt{a_{\rm in}}&=\alpha\,,\label{meanmuin}\\
\Sigma_{\rm in}^2(s)
&=\langle\Delta a_{\rm in}^\dagger\Delta a_{\rm in}\rangle+\frac{1-s}{2}
=\frac{1-s}{2}
\label{varmuin}
\end{align}
(here and throughout we use $\Delta O=O - \expt{O}$).  The three
canonical quasidistributions~\cite{GarrisonChiao2008} correspond to $s=+1$
(normal ordering) for the $P$-function, $s=0$ (symmetric ordering) for the
Wigner~$W$-function, and $s=-1$ (antinormal ordering) for the Husimi~$Q$-distribution.

The mean complex amplitude and $s$-ordered variance of the output state are~\cite{ComCav13}
\begin{align}
\expt{a_{\rm out}}&=G\expt{a_{\rm in}}\,, \label{meanmu}\\
\begin{split}
\Sigma_{\rm out}^2(s)
&=\langle\Delta a_{\rm out}^\dagger\Delta a_{\rm out}\rangle+\frac{1-s}{2}\\
&=\mu^2(G^2-1)+\frac{1-s}{2}\,. \label{varmu}
\end{split}
\end{align}
The amplified input noise is $G^2(1-s)/2$, so the $s$-ordered noise added by the
amplification is $\Sigma_{\rm out}^2(s)-G^2(1-s)/2=(G^2-1)[\mu^2-(1-s)/2]$.  Referred
to the input, this added noise becomes
\begin{align}
{\mathcal A}(s)\equiv \frac{\Sigma_{\rm out}^2(s)}{G^2}-\frac{1-s}{2}
=\bigg(1-\frac{1}{G^2}\bigg)\bigg(\mu^2-\frac{1-s}{2}\bigg)\,.
\end{align}
Following~\cite{CavComJiaPan12}, we prefer to deal with an added-noise
number that has all the gain dependence removed,
\begin{align}
A(s) \equiv \frac{{\mathcal A}(s)}{1-1/G^2}
=\mu^2-\frac{1-s}{2}\,,
\end{align}
which can be thought of as the added noise in the high-gain limit.  Notice that both ${\mathcal A}(s)$ and
$A(s)$ are zero for $2\mu^2=1-s$ and negative for $2\mu^2<1-s$.

How one thinks about amplifier noise depends on the operator ordering one adopts. The traditional way to think about amplifier noise is in terms of symmetric ordering ($s=0$).  Then an ideal linear amplifier adds half a quantum of noise, a perfect amplifier adds no noise, and an immaculate linear amplifier subtracts half a quantum of noise.  For comparing measurements of quadrature components at the input and output, it is more informative to think in terms of
antinormal ordering ($s=-1$), in which case an ideal linear amplifier adds no noise, and perfect and immaculate amplifiers subtract half a quantum and a full quantum of noise, respectively.  For normal ordering ($s=+1$), there is no input noise and so no amplified input noise; all the output noise is added noise, with an ideal linear amplifier, a perfect amplifier, and an immaculate amplifier adding a full quantum, a half a quantum, and no noise, respectively.  This behavior is illustrated in Fig.~\ref{fig2}.

It is worth noting that the perspective of normal ordering is what inspired Ralph and Lund's original terminology, ``noiseless linear amplifier,'' for what is here called an immaculate amplifier.  The ordering dependence of what one means by ``noiseless'' prompts us in this paper to characterize  the nondeterministic $\mu^2$-amplifiers as having \emph{reduced noise}, relative to the ideal linear amplifier, and to refer to particular cases as immaculate and perfect.

\begin{figure}[htbp]
\includegraphics[width=0.45\textwidth]{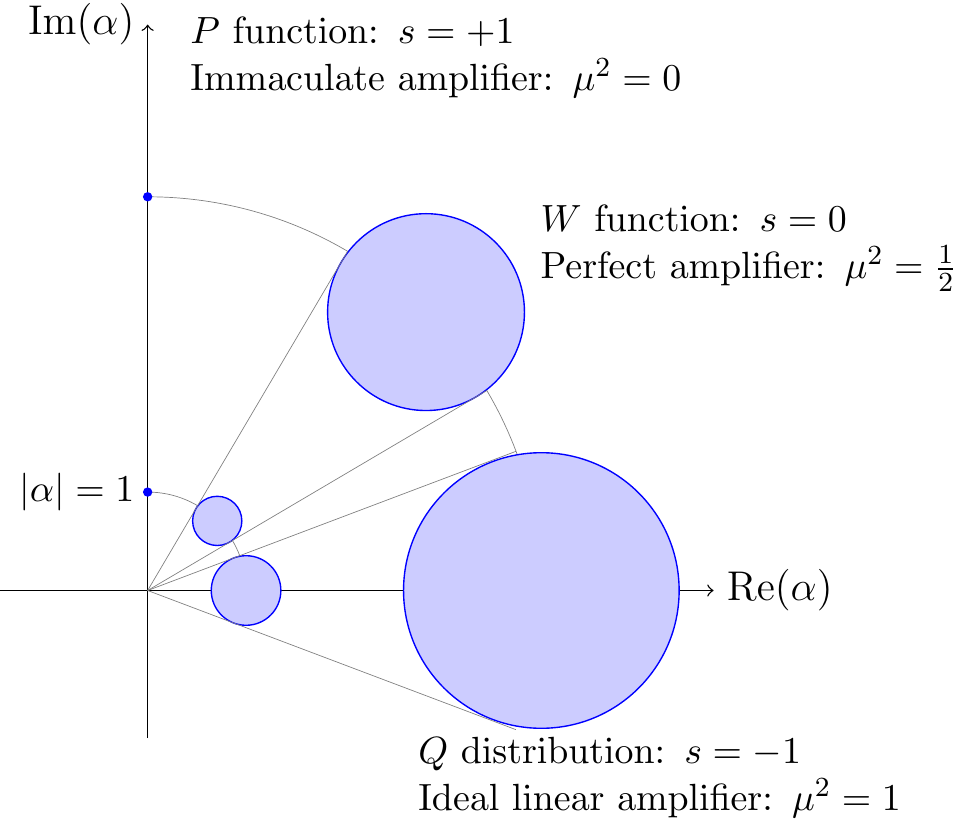}
\caption{(Color online).  The output noise of a $\mu^2$-amplifier looks like a rescaled version of the input noise when $s=1-2\mu^2$; i.e., the $s$-ordered quasiprobability distributions are matched to the output noise of the amplifier, and there is apparently no added noise at the output.  This is depicted in the figure for an immaculate amplifier ($\mu^2=0$, $s=1$, $P$-function), a perfect amplifier ($\mu^2=1/2$, $s=0$, $W$-function), and the ideal linear amplifier ($\mu^2=1$, $s=-1$, $Q$-distribution).  For the immaculate amplifier, both the input and outputs are coherent states, with $P$-functions given by $\delta$ functions and thus no noise as measured by normally ordered moments.  For the other two cases, the grey lines show that the input noise is rescaled by the gain to give the output noise.  The rescaling holds even for nonideal linear amplifiers ($\mu^2>1$, $s<-1$), but that situation is not illustrated in the figure.}\label{fig2}
\end{figure}

Since fidelity, rather than added noise, is the favored way of characterizing the performance
of these amplifiers, we note that the fidelity of the output state~(\ref{eq:E}) with
the target state $|G\alpha\rangle$ is
\begin{align}\label{eq:Fmu1}
F(\mu^2)=\bra{G\alpha}\rho_\textrm{out}\ket{G\alpha}=
\frac{1}{\mu^2(G^2-1)+1}\,.
\end{align}

The unphysical amplifiers for $\mu^2<1$ introduced in this section are purely mathematical constructs.
To be physical, such amplifiers must be probabilistic.  We can take a step toward physicality by
noting that Ref.~\cite{ComCav13} formulated an argument based on the uncertainty principle, which concluded that the success probability of a $\mu^2$-amplifier is bounded by
\begin{align}
\Pcheck(\mu^2)
\le\frac{\mu^2(G^2-1)+1}{G^2}\,.
\label{eq:Pmu}
\end{align}
Alternative state-discrimination based arguments that give this bound were originally given for $\mu^2=0$ in~\cite{RalLun09}; these arguments were later strengthened in~\cite{ComCav13}.  Not surprisingly, this bound is only a restriction on the success probability for $\mu^2<1$, where it can be conveniently re-expressed as a bound on the probability-fidelity product, $\Pcheck(\mu^2)F(\mu^2)\le 1/G^2$, which is independent of $\mu^2$.  In order to have a gain-independent
measure of performance in the following, we work with the quantity
$G^2\Pcheck(\mu^2)F(\mu^2)$, calling this gain-corrected quantity the
probability-fidelity product (PFP). It satisfies the bound
\begin{equation}
G^2\Pcheck(\mu^2)F(\mu^2)\le 1\,,
\label{eq:PFPapproxbound}
\end{equation}
which we adopt as a convenient benchmark for evaluating amplifier performance.

The physical amplifiers for $\mu^2\ge1$ satisfy the bound~(\ref{eq:PFPapproxbound}), with equality
attained only by the ideal linear amplifier.  For $\mu^2<1$, cloning and state-discrimination
arguments~\cite{ComCav13} suggest that the bound~(\ref{eq:PFPapproxbound}) is an absolute bound
on performance.  In Sec.~\ref{sec:exact_n_equals_1}, however, we show that physical $\mu^2$-amplifiers
that have the smallest high-fidelity operating region can violate this bound, and we
investigate there the consequences of this violation and its implications for the
benchmark~(\ref{eq:PFPapproxbound}).

\section{Constructing $\mu^2$--amplifiers}\label{subsec:immac}
Construction of a family of physical $\mu^2$-amplifiers is quite simple.
First we perform physical immaculate amplification with gain $g_1$ and success
probability $\Pcheck$.  Conditional on success of this immaculate amplification,
we then perform (possibly) nonideal amplification with gain $g_2$ and mean
number of quanta $\bar n$.  The combined action of these steps results in a
$\mu^2$-amplifier with gain $G=g_1 g_2$ and success probability
$p_\checkmark$.  The amplifier only works effectively within a disk of
input states satisfying $|\alpha|\ll \sqrt{N}/g_1$, where $N$ is a number-basis
cutoff.

To see how the parameters are related, we first construct unphysical versions
of this scenario using the unphysical immaculate amplifier of Sec.~\ref{sec:laprior};
this shows that the concatenated version of $\mu^2$-amplifiers is really no different from
the $\mu^2$-amplifiers of Sec.~\ref{sec:laprior}.  We then turn in Sec.~\ref{sec:phys_realization}
to constructing physical versions using a model drawn from Ref.~\cite{ComCav13}.

\subsection{Nonphysical construction}
\label{sec:nonphys}

Consider first the physical and unphysical amplifiers of Sec.~\ref{sec:laprior}.  We cascade
the unphysical immaculate amplifier that has gain $g_1$ (stage~1) with a following tunable-noise,
possibly nonideal amplifier with gain $g_2$ and noise $\bar n$ (stage~2).  When acting on
a coherent state $|\alpha\rangle$, the output of stage~1 is the coherent state
$|g_1\alpha\rangle$, which has mean complex amplitude
\begin{align}
\expt{a_{\rm out \,1}}=g_1\expt{a_{\rm in}}=g_1\alpha
\label{mean1}
\end{align}
and $s$-ordered variance
\begin{align}
\Sigma_{\rm out\, 1} ^2(s)=\frac{1-s}{2}\,.
\end{align}

Conditional on the success of stage~1, we follow in stage~2 with nonideal amplification,
which we model by using an ancillary state of form~(\ref{eq:sigmamu}), specified by
$\bar n=\mu^2-1$.  After stage~2, the mean value of the field has undergone the
transformation
\begin{align}
\expt{a_{\rm out \,2}}=g_2g_1 \expt{a_{\rm in\, 1}}\,;
\label{meancascade}
\end{align}
the corresponding $s$-ordered output variance after the second stage is
\begin{align}
\Sigma_{\rm out\, 2}^2(s)=(\bar n+1)(g_2^2 - 1)+\frac{1-s}{2}\,,
\label{varcascade}
\end{align}
That this concatenation yields a $\mu^2$-amplifier can be seen by equating
Eqs.~(\ref{meancascade}) and~(\ref{varcascade}) to Eqs.~(\ref{meanmu}) and~(\ref{varmu}),
which gives
\begin{align}
G&=g_2 g_1\,,\\
\mu^2 & = (\bar n+1)\frac{g_2^2-1}{G^2-1}\,.
\end{align}

More useful are equivalent expressions that are aimed directly at design of a $\mu^2$-amplifier
with gain $G$:
\begin{subequations}
\begin{align}
g_1    & = \frac{G}{g_2},\\
g_2^2& 
=\frac{\mu^2}{\bar n +1}(G^2-1)+1\,.\label{firstg1g2}
\end{align}
\end{subequations}

For $\mu^2\ge1$, we can choose $g_1=1$ ($g_2=G$) and thus $\mu^2=\bar n+1$; this
gives a physical, nonideal linear amplifier. Notice that we could retain the same values of $\mu^2$ and $G^2$ by making $g_1>1$, while maintaining $g_2=G/g_1\ge1$, and increasing $\bar n$, but according to the discussion in Sec.~\ref{sec:phys_realization}, the resulting amplifier would have a subunity success probability, making this is a suboptimal choice. It also suggests that nonideal amplifiers can be made ideal by sacrificing determinism.  Since the nonideal linear amplifier is well understood, we do not consider it for the remainder of the paper, specializing instead to $\mu^2\le1$.

For $\mu^2<1$, it is clear that we must have $g_1>1$ ($g_2<G$); i.e., the
immaculate amplifier of stage~1 must make a contribution to the gain.  It is
useful to note that the fidelity of the output state with the target state
$|G\alpha\rangle$ is still given by Eq.~(\ref{eq:Fmu1}):
\begin{align}
\begin{split}
F(\mu^2)&=\bra{G\alpha}\rho_\textrm{out}\ket{G\alpha}\\
&=\frac{1}{(\bar n+1)(g_2^2-1)+1}\\
&=\frac{1}{\mu^2(G^2-1)+1}\,.
\end{split}
\label{eq:Fmu2}
\end{align}
Since we have not yet put in a physical model of the immaculate amplifier, we
cannot say anything definite about the success probability, except to note
that the uncertainty-principle argument and, hence, the bound~(\ref{eq:Pmu})
still apply.  It should not be surprising, however, that when we put in a
physical model in Sec.~\ref{sec:phys_realization}, we find that to maximize
the success probability, one should make $g_1$ as small as possible, which
means choosing $\bar n=0$.

\subsection{Physical realization}\label{sec:phys_realization}

To construct physical versions of concatenated $\mu^2$-amplifiers, we use a
{\em special case} of the optimal Kraus operators for immaculate amplification,
which were derived in Ref.~\cite{ComCav13}.  An optimal immaculate amplifier
is described by an amplifier map that has a single Kraus operator
\begin{align}
K_\checkmark=P_N\frac{g_1^{a^\dagger a}}{g_1^N}\,,
\label{eq:Kk2}
\end{align}
where $P_N$ is the projector onto the subspace $S_N$ spanned by the first
$N+1$ number states.  The projector $P_N$ enforces a cutoff in the number
basis, which we refer to as the number cutoff; this cutoff means that the
amplifier works effectively only within the operating region $|\alpha|\ll \sqrt{N}/g_1$.

When the Kraus operator~(\ref{eq:Kk2}) acts on a coherent state $\ket\alpha$,
the success probability and fidelity to the target state $\ket{g_1\alpha}$
are given exactly by
\begin{align}
\label{eq:pkopt1}
\Pcheck
&=\bra{\alpha}K_\checkmark^\dagger K_\checkmark\ket{\alpha}
=\frac{e^{-|\alpha|^2}}{g_1^{2N}}e_N\big(g_1^2|\alpha|^2\big)\,,\\
\label{eq:Fopt2}
F
&=\frac{\big|\bra{g_1\alpha}K_\checkmark\ket{\alpha}\big|^2}{\Pcheck}
=e^{-g_1^2|\alpha|^2}e_N\big(g_1^2|\alpha|^2\big)\,,
\end{align}
where
\begin{align}\label{eq:eN}
e_N(x)=\sum_{n=0}^N\frac{x^n}{n!}
\end{align}
denotes the first $N+1$ terms in the expansion of the exponential function.

Within the operating region, $|\alpha|\ll \sqrt{N}/g_1$, the Kraus
operator very nearly maps an input coherent state $|\alpha\rangle$ to the
target state $|g_1\alpha\rangle$, and the fidelity to the target~is
\begin{align}
F&\simeq 1-e^{-g_1^2|\alpha|^2}\left(\frac{eg_1^2|\alpha|^2}{N+1}\right)^{N+1}\,.
\end{align}
Thus, within the operating region, we can regard this model as being an
immaculate amplifier with success probability given approximately by
\begin{align}
\Pcheck \simeq \frac{e^{-|\alpha|^2}}{g_1^{2N}} \le \frac{1}{g_1^{2N}} \,.\label{pcheckapprox}
\end{align}

The output state of the first stage is fed into a nonideal linear amplifier.
Within the operating region, the output state of the first stage is very nearly
the coherent state $\ket{g_1\alpha}$, so Eq.~(\ref{firstg1g2}) applies to the design
of the $\mu^2$-amplifier, and the overall fidelity is given by Eq.~(\ref{eq:Fmu2}).
To maximize the success probability~(\ref{pcheckapprox}) for $\mu^2<1$, it is clear
that we should minimize $g_1$, i.e., maximize $g_2$, and that means choosing $\bar n=0$
for the second stage of the amplification.  Thus, for $\mu^2\le1$, the design
principle~(\ref{firstg1g2}) becomes
\begin{align}
g_2^2=\frac{G^2}{g_1^2}
=\mu^2(G^2-1)+1\,.\label{g1g2}
\end{align}
Within the operating region, the success probability is given by Eq.~(\ref{pcheckapprox}),
and the overall fidelity approximately by (this holds exactly only at $\alpha=0$)
\begin{align}\label{eq:Fmu3}
F(\mu^2)=\frac{1}{\mu^2(G^2-1)+1}=\frac{g_1^2}{G^2}=\frac{1}{g_2^2}\,.
\end{align}
A physical $\mu^2$-amplifier with $\mu^2<1$ is thus a physical immaculate amplifier followed
by an ideal amplifier.

Within the high-fidelity operating region, a physical immaculate ($\mu^2=0$) linear amplifier
has PFP
\begin{align}
G^2\Pcheck(0)F(0)=\frac{1}{G^{2(N-1)}}\,,
\end{align}
which always satisfies the bound~(\ref{eq:PFPapproxbound}).  Indeed, for $N>1$ and any reasonably large gain, the PFP is much smaller than 1.  The conclusion of Ref.~\cite{ComCav13}
was that the optimal immaculate amplifier generally operates far from the bound~(\ref{eq:PFPapproxbound}).

We can now generalize that conclusion to the entire class of physical $\mu^2$-amplifiers for
$\mu^2<1$.  Within the high-fidelity operating region, we can use Eqs.~(\ref{pcheckapprox}) and Eq.~(\ref{g1g2}) to write
\begin{align}\label{eq:pcheck_mu_approx}
\Pcheck(\mu^2)=\frac{1}{g_1^{2N}}
=\frac{\big[\mu^2(G^2-1)+1\big]^N}{G^{2N}}\,.
\end{align}
The resulting PFP,
\begin{align}\label{eq:pfprod_mu_amp_quant_limit}
G^2\Pcheck(\mu^2)F(\mu^2)
=\frac{1}{g_1^{2(N-1)}}
=\frac{\big[\mu^2(G^2-1)+1\big]^{N-1}}{G^{2(N-1)}}\,,
\end{align}
always satisfies the bound~(\ref{eq:PFPapproxbound}).

\section{Bounds on physical $\mu^2$-amplifiers}\label{subsec:upbounds}

The PFP~(\ref{eq:pfprod_mu_amp_quant_limit}) is the central result
of this paper.  It holds approximately within the high-fidelity operating region, $|\alpha|\ll \sqrt{N}/g_1$, but is a strict equality only in the limit $|\alpha|\rightarrow0$.
As noted earlier, cloning and state-discrimination arguments~\cite{ComCav13} suggest that the bound~(\ref{eq:PFPapproxbound}) is an absolute bound on performance.  That Eq.~(\ref{eq:pfprod_mu_amp_quant_limit}) has $G^2\Pcheck(\mu^2)F(\mu^2)=1$ for $N=1$, for all values of $\mu^2$, suggests that the approximations that lead to Eq.~(\ref{eq:pfprod_mu_amp_quant_limit}) need to be re-examined in the case $N=1$.  Indeed, we can calculate exact PFPs for our model of immaculate amplification, and these show that $N=1$ physical $\mu^2$-amplifiers violate the bound~(\ref{eq:PFPapproxbound}).  We consider these exact results and their implications in Sec.~\ref{sec:exact_n_equals_1}.

In this section we explore the consequences of the probability-fidelity product~(\ref{eq:pfprod_mu_amp_quant_limit}) and related results for signal-to-noise ratios and noise figures; thus in this section, we are assuming operation in the
high-fidelity operating region, $|\alpha|\ll \sqrt{N}/g_1$ (strictly speaking, $|\alpha|\rightarrow0$), and we assume $N\ge2$, deferring consideration of $N=1$ to the consideration of exact results in Sec.~\ref{sec:exact_n_equals_1}.

\subsection{Two regimes of operation}
\label{sec:newbounds}

The $\mu^2$-amplifiers are characterized by two parameters, $\mu^2$ and the squared overall gain $G^2$.  To understand the performance of physical $\mu^2$-amplifiers, it is useful to distinguish two quite different regimes in the two-dimensional space of $\mu^2$ and $G^2$ (see Fig.~\ref{fig3}).  The boundary between these two regimes is the line $\mu^2G^2=1$.  Below the boundary line, i.e., $\mu^2G^2<1$, we may regard the immaculate linear amplifier as predominating in the operation of the device, so we call this the {\it immaculate-dominant\/} regime of operation;  above the boundary line, i.e., $\mu^2G^2>1$, we may regard the ideal amplifier as predominating, so we call this the {\it ideal-dominant\/} regime.

\begin{figure}[htbp]
\includegraphics[width=0.5\textwidth]{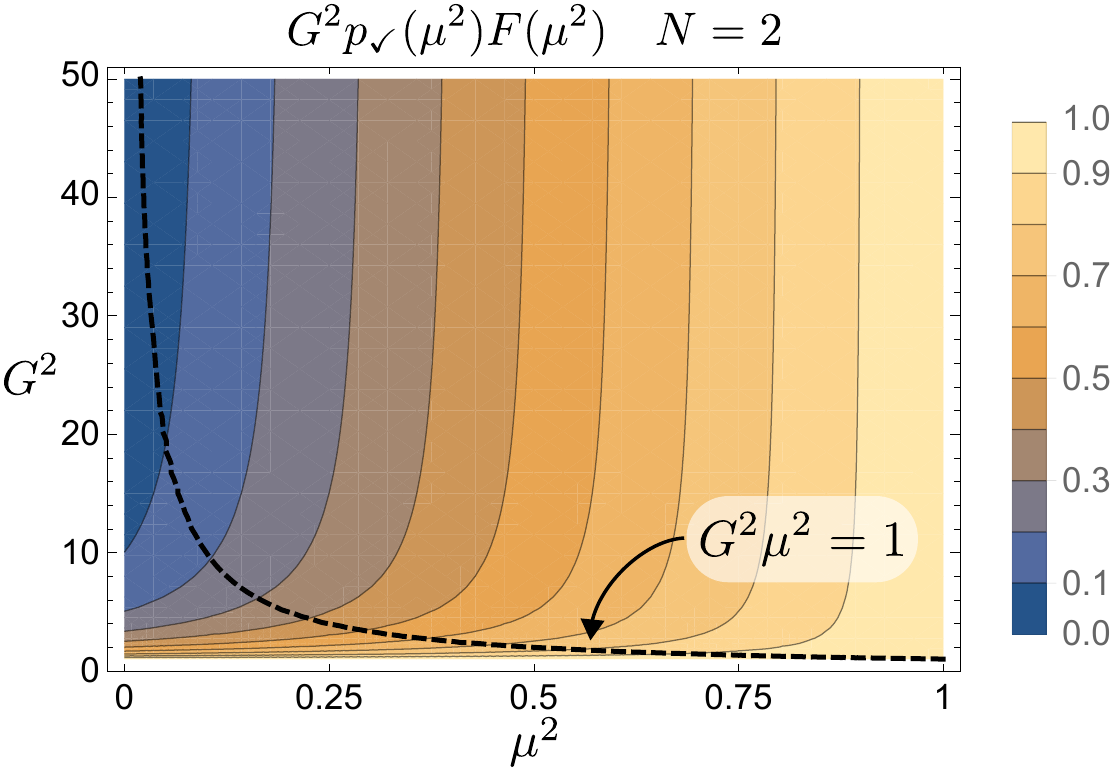}
\caption{(Color online). Contours of the PFP $G^2\Pcheck(\mu^2)F(\mu^2)$ as
a function of $\mu^2$ and $G^2$ for $N=2$.  An ideal linear amplifier ($\mu^2=1$) achieves the maximum, $G^2\Pcheck(\mu^2)F(\mu^2)=1$, for all gains.  The dashed (black) line is the boundary, $G^2\mu^2=1$, between the immaculate-dominant regime of operation, below and to the left of the bounding line, and the ideal-dominant regime, above and to the right of the boundary.  In the extreme ideal-dominant regime, i.e., well above and to the right of the bounding line, the PFP is independent of $G^2$ and given by $\mu^2$; the contours become vertical lines.  In the extreme immaculate-dominant regime, the PFP is independent of $\mu^2$ and given by $1/G^2$.  This behavior is seen in the nearly horizontal contours just above the horizontal axis and in the values on the contours as they contact the vertical axis. }\label{fig3}
\end{figure}

In the extreme immaculate-dominant regime, well below and to the left of the boundary, where $\mu^2\le\mu^2G^2\ll N$, we have
\begin{align}
\begin{split}
&\frac{g_1^2}{G^2}=\frac{1}{g_2^2}=F\simeq 1\,,\\
&\Pcheck\simeq\frac{1}{G^{2N}}\,,\quad
G^2\Pcheck F\simeq\frac{1}{G^{2(N-1)}}\,.
\end{split}
\end{align}
It is notable that in this extreme regime, both the success probability and the PFP are independent of $\mu^2$.

\begin{table*}[t!]
\begin{tabular}{|l|c|c l|c l|}
\hline
& \,$\mu^2=0$ \,
& \multicolumn{2}{c|}{\,$\mu^2=\half$\,}
& \multicolumn{2}{c|}{\,$\mu^2=1$\,}\\
& \,(immaculate)\,
& \multicolumn{2}{c|}{\,(perfect)\,}
& \multicolumn{2}{c|}{\,(ideal)\,}\\\hline
  $\vphantom{\Bigg(}g_1^2$
& $G^2$
& $\displaystyle{\frac{2G^2}{G^2+1}}$
& $(2)$
& $1$
& \\ \hline
  $\vphantom{\Bigg(}g_2^2$
& $1$
& $\displaystyle{\frac{G^2+1}{2}}$
& $\textstyle{(G^2/2)}$
& $G^2$
& \\ \hline
  $\vphantom{\Bigg(}F$
& $1$
& $\displaystyle{\frac{2}{G^2+1}}$
& $(2/G^2)$
& $\displaystyle{\frac{1}{G^2}}$
& \\ \hline
  $\vphantom{\Bigg(}\Pcheck$
& $\displaystyle{\frac{1}{G^{2N}}}$
& $\displaystyle{\frac{1}{2^N}\bigg(1+\frac{1}{G^2}\bigg)^N}$
& $(1/2^N)$
& $1$
& \\ \hline
  $\vphantom{\Bigg(}G^2\Pcheck F$
& $\displaystyle{\frac{1}{G^{2(N-1)}}}$
& $\displaystyle{\frac{1}{2^{N-1}}\bigg(1+\frac{1}{G^2}\bigg)^{N-1}}$
& $(1/2^{N-1})$
& $1$
& \\ \hline
$\vphantom{\Bigg(}{\rm NF}(-1)$
& $G^{2(N-1)}$
& $\displaystyle{{2^{N-1}}\bigg(1+\frac{1}{G^2}\bigg)^{-(N-1)}}$
& $(2^{N-1})$
& $1$
& \\ \hline
$\vphantom{\Bigg(}{\rm NF}(0)$
& $G^{2(N-1)}$
& $\displaystyle{{2^N}\bigg(1+\frac{1}{G^2}\bigg)^{-N}}$
& $(2^N)$
& $\displaystyle{2-\frac{1}{G^2}}$
& $(2)$ \\ \hline
\end{tabular}
\caption{Summary of properties of immaculate, perfect, and ideal linear amplifiers, when operating
in the high-fidelity operating region.  Results in parentheses are for the high-gain limit, $G^2\gg1$.
For any gain, an immaculate amplifier operates in the  immaculate-dominant regime, and an ideal amplifier
operates in the ideal-dominant regime.  A perfect amplifier passes between the two regimes at gain $G^2=2$;
for high gain, $G^2\gg1$, a perfect amplifier operates in the extreme ideal-dominant regime. The noise
figure used here includes the correction for success probability, as defined in Eq.~(\ref{eq:NFs})}.
\label{tab:Table}
\end{table*}

In the extreme ideal-dominant regime, far above the bounding line, where $G^2\ge\mu^2G^2\gg N$, we have
\begin{align}
\begin{split}
&\frac{g_1^2}{G^2}=\frac{1}{g_2^2}=F\simeq\frac{1}{\mu^2G^2}\;,\\
&\Pcheck\simeq\mu^{2N}\,,\quad
G^2\Pcheck F\simeq\mu^{2(N-1)}\,.
\end{split}
\end{align}
It is notable that in this extreme regime, both the success probability and the PFP are independent of gain.

Figure~\ref{fig3} illustrates the behavior of the PFP by plotting
its contours as a function of $\mu^2$ and $G^2$ for $N=2$, which is representative of all the cases $N\ge2$.  The features of the plot become sharper as $N$ increases from 2.

Table~\ref{tab:Table} summarizes the properties of the three special amplifiers introduced in Sec.~\ref{sec:laprior}.

\subsection{$s$-ordered root-probability--SNR product and noise figure}

Bounds on the {\em antinormally ordered\/} root-probability--SNR (signal-to-noise ratio)
product, $\sqrt{\Pcheck}{\rm SNR}$, were also obtained in Ref.~\cite{ComCav13}.  SNRs are a particular measure of the distinguishability of quantum states, cast in terms of the ability to
resolve a signal within its associated noise.  The root-probability--SNR product is the signal-to-noise ratio corrected for the fact that the success probability $\Pcheck$ reduces the number of chances at the output of the amplifier to obtain information about a signal.

Assuming an input coherent state with real $\alpha$, the quadrature components $x_1$ and $x_2$ of Eq.~(\ref{eq:x1x2}) represent the amplitude and phase quadratures.  The antinormally ordered signal-to-noise ratio is defined as ${\rm SNR}\equiv\langle x_1\rangle/\Delta x_1=\langle x_1\rangle/\Delta x_2$, where $\Delta x_1=\Delta x_2$ are the square roots of the antinormally ordered variances in $x_1$ and $x_2$.  The uncertainty-principle bound~(\ref{eq:Pmu}) on success probability is equivalent to the requirement that amplification not increase the SNR measure of resolvability, i.e.,
\begin{align}
\sqrt{\Pcheck}\,{\rm SNR}_{\rm out}\le{\rm SNR}_{\rm in}=\sqrt2\alpha\,.
\label{eq:SNR_bound}
\end{align}
The optimal immaculate amplifier did not come close to saturating this bound.

We can generalize the SNR considerations to the arbitrary operator orderings considered in Sec.~\ref{sec:laprior} by using $s$-ordered variances in the definition of the~\hbox{SNR}.
The $s$-ordered input SNR is
\begin{align}
{\rm SNR}_{\rm in}(s)=\frac{\sqrt2\alpha}{\sqrt{(1-s)/2}}\,.
\label{eq:SNR_in}
\end{align}
From Eqs.~(\ref{meanmu}) and~(\ref{varmu}), we determine that the $s$-ordered output SNR is
\begin{align}
{\rm SNR}_{\rm out}(s)=\frac{\sqrt2 G\alpha}{\sqrt{\mu^2(G^2-1)+(1-s)/2}}\,.
\label{eq:SNR_out}
\end{align}

It is a good idea to pause here to consider these SNRs for the ideal linear amplifier, i.e., $\mu^2=1$, and for various operator orderings.  For antinormal ordering ($s=-1$), ${\rm SNR}_{\rm out} = {\rm SNR}_{\rm in}$  irrespective of gain~\cite{CavComJiaPan12}, so there is no degradation of SNR. The lack of degradation is a manifestation of preservation of signal-to-noise for simultaneous  measurement of both quadratures (i.e., heterodyne measurement).  For symmetric ordering ($s=0$), ${\rm SNR}_{\rm out} = {\rm SNR}_{\rm in}/\sqrt{2}$ in the high-gain limit. This is the traditional view of an amplifier in which SNR is degraded even by ideal linear amplification.  Normal ordering
results in a singular ${\rm SNR}_{\rm in}$, so we do not consider it.

When the amplification is probabilistic, ${\rm SNR}_{\rm out}$ is not the relevant measure of overall performance at the output.  Instead, the root-probability--SNR product, $\sqrt{ \Pcheck} {\rm SNR}_{\rm out}(s)$ is the relevant measure, with the success probability given by Eq.~(\ref{eq:pcheck_mu_approx}).  The root-probability--SNR product is the right measure because it accounts for the reduced chance of measuring an output signal.

Instead of looking at $s$-ordered input and output SNRs, however, it is more informative to look
at the $s$-ordered noise figure, which is the input-to-output ratio of the appropriate squared SNRs,
\begin{align}\label{eq:NFs}
\begin{split}
{\rm NF}(s)
&=\frac{{\rm SNR}_{\rm in}^2(s)}{\Pcheck{\rm SNR}_{\rm out}^2(s)}\\
&=\frac{G^{2(N-1)}}{(1-s)/2}
\frac{\mu^2(G^2-1)+(1-s)/2}{\big[\mu^2(G^2-1)+1\big]^N}\,.
\end{split}
\end{align}
For antinormal ordering ($s=-1$), the noise figure is the inverse of the PFP~(\ref{eq:pfprod_mu_amp_quant_limit}), so our discussion of the two regimes of
operation in Sec.~\ref{sec:newbounds} can be applied directly to ${\rm NF}(-1)$.  In the extreme immaculate-dominant and ideal-dominant regimes of operation, the antinormally ordered noise figure becomes
\begin{align}
{\rm NF}(-1)\simeq
\begin{cases}
G^{2(N-1)}\,,&\mbox{for $N\mu^2G^2\ll1$},\\
\mu^{-2(N-1)}\,,&\mbox{for $\mu^2G^2\gg N$},
\end{cases}
\end{align}
Normal ordering ($s=+1$) gives a singular noise figure, so we do not consider it here.
Table \ref{tab:Table} summarizes the noise figure for the three special $\mu^2$-amplifiers.
The unsurprising conclusion is that the ideal linear amplifier is the best with respect to
this measure.

\section{Exact results and the $N=1$ case}\label{sec:exact_n_equals_1}

\subsection{Exact results}

The analysis in the previous two sections is close to exact when the physical $\mu^2$-amplifier
operates in the high-fidelity operating region,
\begin{align}
|\alpha|\ll \sqrt{N}/g_1\equiv|\tilde\alpha|\,.
\end{align}
In this section we examine the case when $N$ is small---in particular we focus on $N\in \{1,2\}$---so that the operating region is a very small disk at the origin, but we do not restrict the input amplitude $|\alpha|$ to this high-fidelity region.  As we have done since the end of Sec.~\ref{sec:nonphys}, we make the second stage of our $\mu^2$-amplifiers an ideal amplifier, leaving aside the possibility of a nonideal second stage.

Let's return now to the construction of physical $\mu^2$-amplifiers as in Sec.~\ref{sec:phys_realization}.  In doing so, recall that the fidelity of an arbitrary state, $\rho$, with some target coherent state, $\ket{\alpha}$, is the Husimi $Q$-function evaluated at $\ket{\alpha}$, namely, $F(\rho,\ket{\alpha}) = \bra{\alpha}\rho\ket{\alpha} = \pi Q_\rho(\alpha)$.  Thus the fidelity between the output of the $\mu^2$-amplified state and the target coherent state $G\alpha$ is given by,
\begin{align}
F&= \bra{G\alpha}\rho_{\mathrm{out}}\ket{G\alpha}
= \pi Q_{\rho_{\mathrm{out}}}(G\alpha)\,.
\end{align}

Now denote the state after the initial immaculate amplification with gain $g_1$ by
$\rho'=K_\checkmark\ket{\alpha}\bra{\alpha}K_\checkmark^\dagger/\Pcheck$.  The fidelity~(\ref{eq:Fopt2}) between $\rho'$ and the coherent state $\ket{g_1\alpha}$ is the $Q$-function
\begin{align}
\pi Q_{\rho'}(g_1\alpha)
=\frac{\big|\bra{g_1\alpha}K_\checkmark\ket{\alpha}\big|^2}{\Pcheck}
=e^{-g_1^2|\alpha|^2}e_N\big(g_1^2|\alpha|^2\big)\,.
\end{align}
State $\rho'$ is fed into an ideal linear amplifier of gain $g_2$.  For an ideal linear amplifier of gain $g$, it was shown in~\cite{CavComJiaPan12} that the $Q$-function transforms as
$Q_{\rho_{\mathrm{out}}}(\beta) = Q_{\rho_{\mathrm{in}}}(\beta/g)/g^2$.  Applying this to our present case, we have
\begin{align}
Q_{\rho_{\mathrm{out}}}(G\alpha) &= \frac{Q_{\rho'}\left( G\alpha/g_2\right)}{g_2^2}
=\frac{g_1^2Q_{\rho'}\left( g_1\alpha\right)}{G^2}\,,
\end{align}
where we use $G=g_1g_2$.  Thus the output fidelity is given by
\begin{align}
F&= \frac{g_1^2e^{-g_1^2|\alpha|^2}}{G^2}e_N\big(g_1^2|\alpha|^2\big)\,.\label{Fexact}
\end{align}
The success probability is that given by Eq.~(\ref{eq:pkopt1}), and the exact result for the
PFP is
\begin{align}\label{eq:PFP}
G^2\Pcheck F=
\frac{e^{-(g_1^2+1)|\alpha|^2}}{g_1^{2(N-1)}}e_N^2\big(g_1^2|\alpha|^2\big)\,.
\end{align}
In the expressions for $F$, $\Pcheck$, and $G^2\Pcheck F$, we can write $g_1^2$ in terms of the primary
parameters, $\mu^2$ and $G^2$, by using the design principle~(\ref{g1g2}).

Before looking at particular examples, let's establish an exact bound on the fidelity-probability product~(\ref{eq:PFP}).  To do so, notice that by using Eqs.~(\ref{eq:pkopt1}) and~(\ref{eq:Fopt2}), we can write Eq.~(\ref{eq:PFP}) as
\begin{align}
G^2\Pcheck F=
g_1^2\big|\bra{g_1\alpha}K_\checkmark\ket{\alpha}\big|^2\;.
\end{align}
Now, following~\cite{ComCav13}, we establish an upper bound on the real quantity $\bra{g_1\alpha}K_\checkmark\ket{\alpha}$:
\begin{align}
\begin{split}
\bra{g_1\alpha}K_\checkmark\ket{\alpha}
&=\sum_{n=0}^N\frac{g_1^n}{g_1^N}\langle g_1\alpha|n\rangle\langle n|\alpha\rangle\\
&\le\sum_{n=0}^\infty\langle g_1\alpha|n\rangle\langle n|\alpha\rangle
=\langle g_1\alpha|\alpha\rangle\,.
\end{split}
\end{align}
Thus we have the exact bound
\begin{align}\label{eq:PFPexactbound}
G^2\Pcheck F
\le g_1^2\big|\langle g_1\alpha|\alpha\rangle\big|^2
=g_1^2e^{-(g_1-1)^2|\alpha|^2}
\equiv{\rm PFP}_0\,.
\end{align}
For an ideal linear amplifier, $g_1=1$, so ${\rm PFP}_0=1$ for all $\alpha$, which duplicates the
bound~(\ref{eq:PFPapproxbound}).  For $g_1>1$, PFP$_0$ decreases from $g_1^2$ at $|\alpha|=0$ to zero as $|\alpha|\rightarrow\infty$, passing through 1 when
\begin{align}
|\alpha|^2=\frac{2\ln g_1}{(g_1-1)^2}\equiv|\alpha_0|^2\,.
\end{align}
For $|\alpha|>|\alpha_0|$ , the exact bound PFP$_0$ is stricter than the approximate bound~(\ref{eq:PFPapproxbound}), but for $|\alpha|<|\alpha_0|$, the exact bound permits violation of the approximate PFP bound.

\begin{figure}[htbp]
\includegraphics[width=0.45\textwidth]{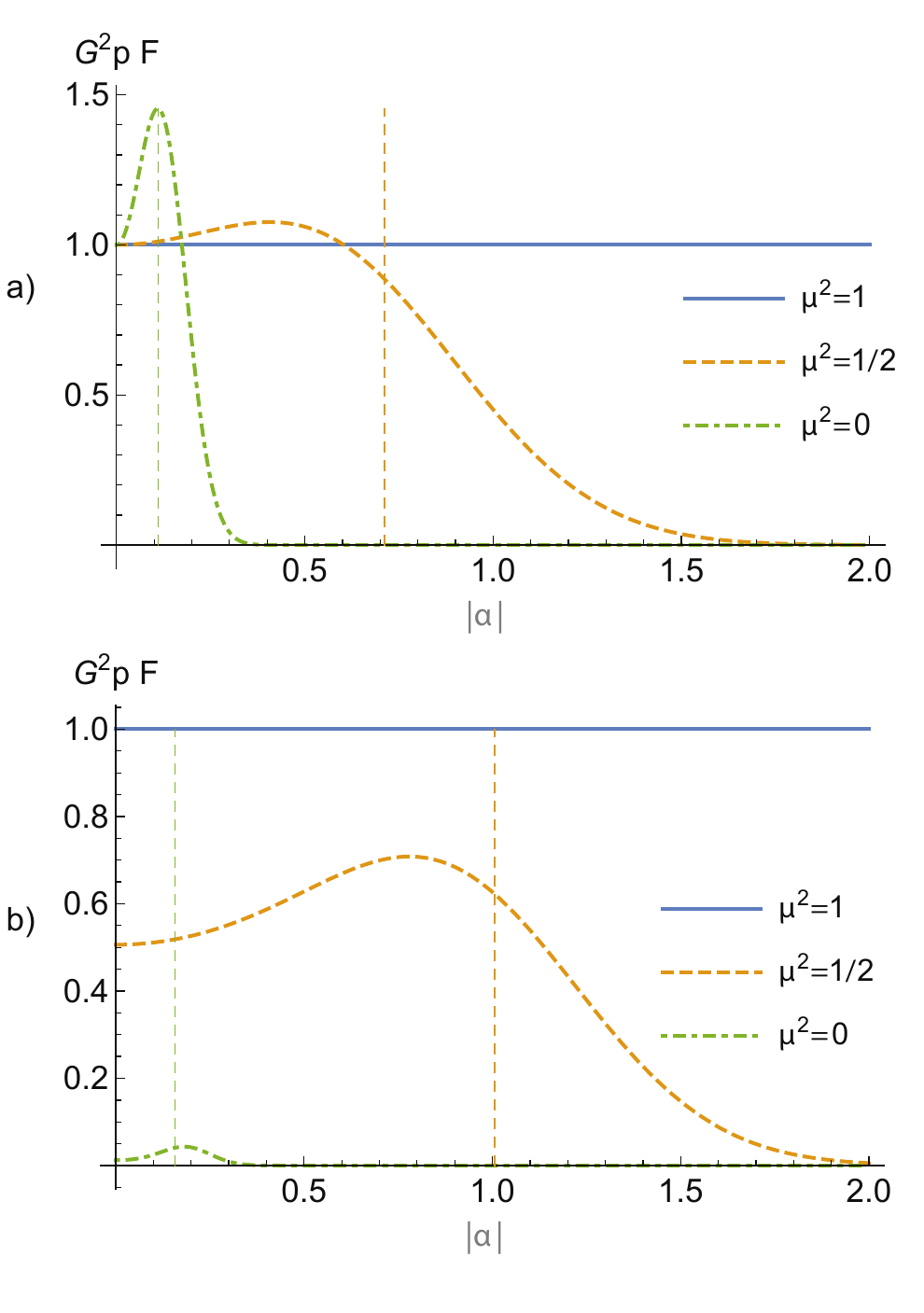}
\caption{(Color online) PFP $G^2\Pcheck(\mu^2)F(\mu^2)$ as a function of input coherent-state amplitude $|\alpha|$ for ideal ($\mu^2=1$), perfect ($\mu^2=\frac12$), and immaculate ($\mu^2=0$) amplifiers
with $G=9$ and (a)~$N=1$, (b)~$N=2$.  The vertical dashed lines are at
$|\tilde\alpha|=\sqrt{N}/g_1$ for $\mu^2=0$ and $\mu^2=1/2$.
}\label{fig4}
\end{figure}

To interpret the exact bound, consider a scenario in which one omits the first, immaculate stage
with gain $g_1$ in our $\mu^2$-amplifiers, but then applies the second, ideal stage with gain $g_2=G/g_1$
to the input state $\ket{\alpha}$.  Let $\rho_{\rm out}$ now denote the output of this scenario.
After the second, ideal stage, the fidelity to the target coherent state $|G\alpha\rangle$~is
\begin{align}
\begin{split}
\big|\bra{G\alpha}\rho_{\rm out}\ket{G\alpha}\big|^2
&=\pi Q_{\rho_{\rm out}}(G\alpha)\\
&=\frac{\pi Q_{\rho_{\rm in}}(G\alpha/g_2)}{g_2^2}\\
&=\frac{g_1^2\big|\langle\alpha|g_1\alpha\rangle\big|^2}{G^2}\,.
\end{split}
\end{align}
Since this scenario can be carried out deterministically, its PFP achieves the upper bound PFP$_0$:
\begin{align}
G^2\big|\bra{G\alpha}\rho_{\rm out}\ket{G\alpha}\big|^2=g_1^2\big|\langle\alpha|g_1\alpha\rangle\big|^2
={\rm PFP}_0\,.
\end{align}
It is worth stressing what this means for the exact bound~(\ref{eq:PFPexactbound}): {\it the PFP for a $\mu^2$-amplifier is never better than the PFP obtained by omitting the immaculate stage of amplification and replacing it with doing nothing.}

\begin{table}[h!]
\begin{tabular}{|l|c|c|c|}
\hline
					& \,$\mu^2=0$ \, & \,$\mu^2=\half$\, & \,$\mu^2=1$ \, \\
					& \,(immaculate)\, & \,(perfect)\, & \, (ideal)\, \\
\hline
$g_1$						& $9$ 		& $1.406$  &  $ 1$  	\\ \hline
$g_2$						& $1$  		& $6.403$  &   $9$ 	\\ \hline
$|\tilde\alpha|=1/g_1$ ($N=1$)	& $0.111$  & $0.711$  &   - \\ \hline
$|\tilde\alpha|=\sqrt2/g_1$ ($N=2$)	& $0.157$  & $1.006$  &   - \\ \hline
\end{tabular}
\caption{Design parameters and input coherent-state amplitude, $|\tilde\alpha|=\sqrt{N}/g_1$, that
 defines the high-fidelity operating region.  In the fourth column the symbol ``-" indicates that the ideal linear amplifier works over the entire phase plane.}\label{tab:TableDesign}
\end{table}

Now we analyze some representative examples to illustrate the features
of the exact analysis.  We focus on the parameters $G=9$, $N\in \{1,2\}$,
and $\mu^2\in \{0,\frac12,1\}$, i.e., the values of $\mu^2$ corresponding to immaculate,
perfect, and ideal amplifiers.  These give the design parameters and high-fidelity
operating region listed in Table~\ref{tab:TableDesign}.

It is useful to record the relevant exact quantities for $N=1$ and $N=2$:
\begin{subequations}
\begin{align}
\mbox{$N=1$:}\hspace{0.5em}
&\Pcheck=\frac{e^{-|\alpha|^2}}{g_1^2}\big(1+g_1^2|\alpha|^2\big)\,,\\
&F= \frac{e^{-g_1^2|\alpha|^2}}{g_2^2}\big(1+g_1^2|\alpha|^2\big)\;,\\
&G^2\Pcheck F=e^{-(g_1^2+1)|\alpha|^2}\big(1+g_1^2|\alpha|^2\big)^2\,,\label{eq:PFPN1}
\end{align}
\end{subequations}
\begin{subequations}
\begin{align}
\mbox{$N=2$:}\hspace{0.5em}
&\Pcheck=\frac{e^{-|\alpha|^2}}{g_1^4}\big(1+g_1^2|\alpha|^2+g_1^4|\alpha|^4/2\big)\,,\\
&F= \frac{e^{-g_1^2|\alpha|^2}}{g_2^2}\big(1+g_1^2|\alpha|^2+g_1^4|\alpha|^4/2\big)\;,\\
&G^2\Pcheck F\nonumber\\
&\quad=\frac{e^{-(g_1^2+1)|\alpha|^2}}{g_1^2}\big(1+g_1^2|\alpha|^2+g_1^4|\alpha|^4/2\big)^2\,.
\end{align}
\end{subequations}

Figure~\ref{fig4} gives plots of the PFP for the three special amplifiers.  When $N=1$,
both the immaculate and perfect amplifiers have a ``bump'' that beats the na\"ive, approximate bound,
$G^2pF\leq 1$, of Eq.~(\ref{eq:PFPapproxbound}).  Exceeding the approximate bound is
most pronounced for a purely immaculate amplifier; generically, as $\mu^2$ increases, the
``bump" becomes smaller and occurs at a higher input amplitude.  For $N=2$ (and all
larger $N$), even though the bump persists, it becomes less pronounced and never beats the
approximate bound, which is achieved for all $\alpha$ by the ideal linear amplifier.  Thus, for
the remainder of this section, we focus on the $N=1$ case.

\begin{figure}[htbp]
\includegraphics[width=0.45\textwidth]{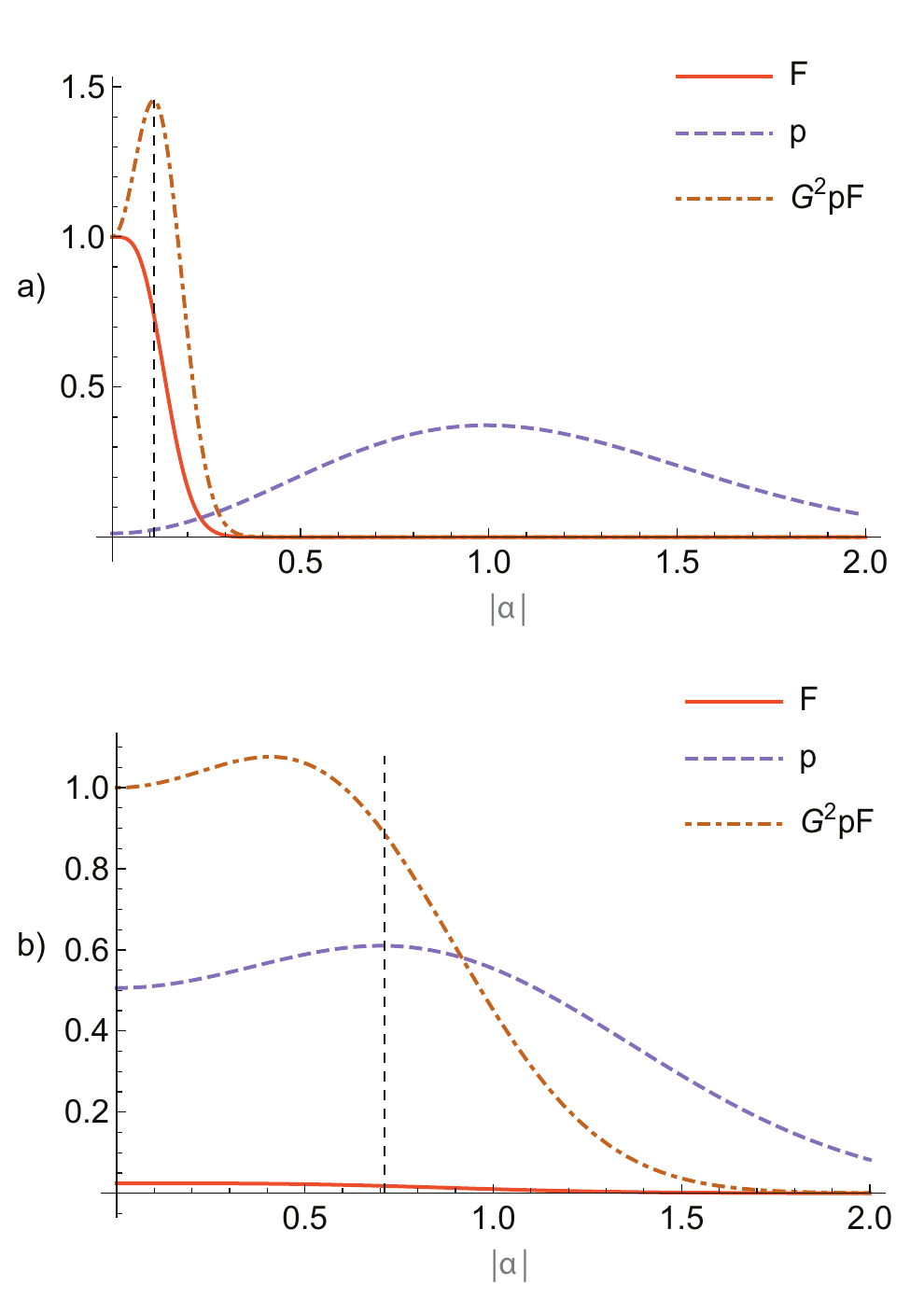}
\caption{(Color online) Fidelity $F(\mu^2)$, success probability $\Pcheck(\mu^2)$, and PFP $G^2\Pcheck(\mu^2)F(\mu^2)$ as functions of input coherent-state amplitude $|\alpha|$ for amplifiers
with $N=1$ and gain $G=9$: (a)~immaculate amplifier ($\mu^2=0$) and (b)~perfect amplifier ($\mu^2=1/2$).
The vertical dashed lines are at $|\tilde\alpha|=1/g_1$.}
\label{fig5}
\end{figure}

It is easy to derive from the $N=1$ PFP~(\ref{eq:PFPN1}) that the peak of the PFP bump
occurs at
\begin{align}\label{eq:bump}
|\alpha_{\rm bump}|^2=\frac{1}{g_1^2}\frac{g_1^2-1}{g_1^2+1}\le\frac{1}{g_1^2}=|\tilde\alpha|^2
\end{align}
and that the value of the PFP at the peak is
\begin{align}\label{eq:bumpvalue}
\big(G^2\Pcheck F\big)_{\rm bump}
=\frac{4}{e}\frac{e^{1/g_1^2}}{(1+1/g_1^2)^2}
\le\frac{4}{e}\,.
\end{align}
For an immaculate amplifier, for which $g_1=G$, the bump peaks just inside $|\tilde\alpha|$
in the high-gain limit and has peak value~$4/e=1.472$.  For a perfect amplifier, rewriting Eqs.~(\ref{eq:bump}) and~(\ref{eq:bumpvalue}) in terms of the overall gain,
\begin{align}
g_1^2|\alpha_{\rm bump}|^2
&=\frac{G^2-1}{3G^2+1}\le\frac13\,,\\
\big(G^2\Pcheck F\big)_{\rm bump}
&=\frac{16}{9\sqrt{e}}\frac{e^{1/2G^2}}{(1+1/3G^2)^2}
\le\frac{16}{9\sqrt{e}}\,,
\end{align}
shows that in the high-gain limit, the bump peaks just inside $|\tilde\alpha|/\sqrt3$ and has
peak value $16/9\sqrt e=1.078$.

\begin{figure}[htbp]
\includegraphics[width=0.45\textwidth]{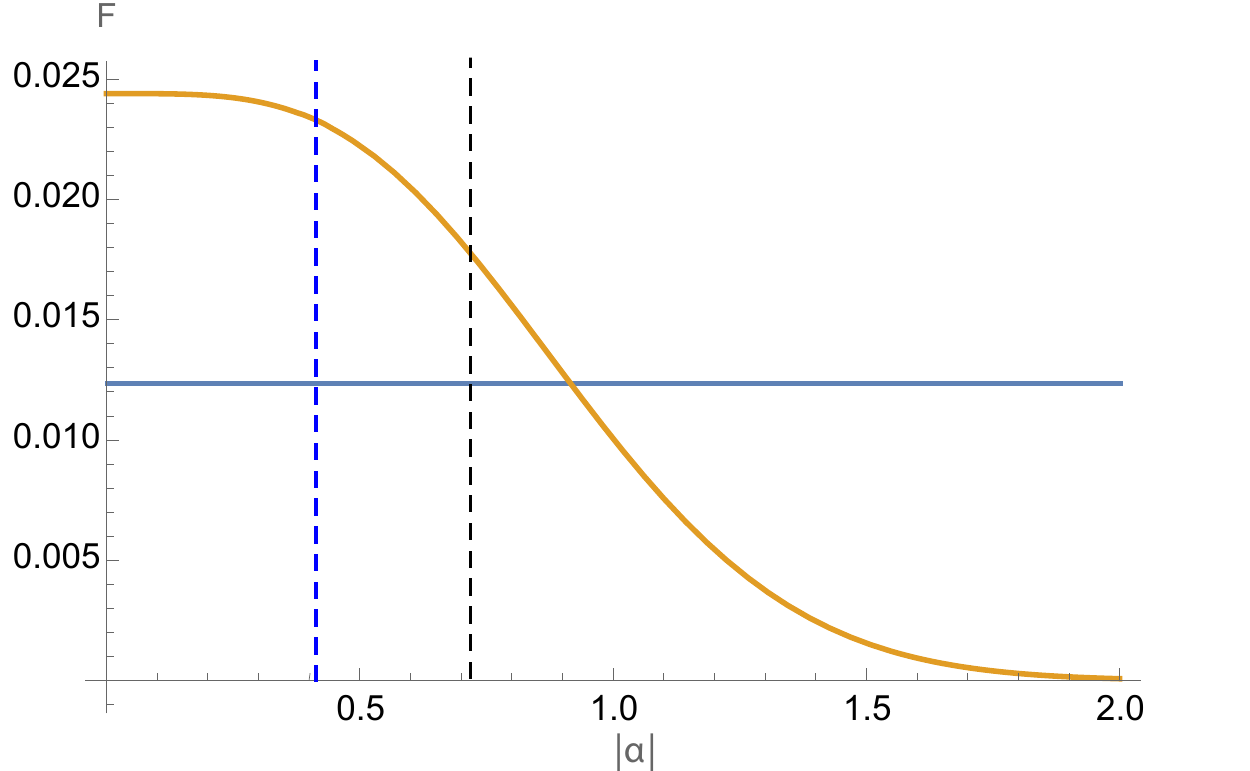}
\caption{(Color online) Output fidelity $F(\mu^2)$ for perfect amplification ($\mu^2=1/2$) with $N=1$ and $G=9$; this is the fidelity of Fig.~\ref{fig5}(b) on an expanded vertical scale.  The black and blue dashed lines are at $|\tilde\alpha|=1/g_1=0.711$ and $|\alpha_{\mathrm{bump}}|=0.573|\tilde\alpha|= 0.407$, respectively.  The fidelity for an ideal amplifier with the same gain is shown for comparison.
}\label{fig6}
\end{figure}

Figure~\ref{fig5} plots the fidelity, the success probability, and the PFP for the $N=1$ immaculate and
perfect amplifiers.  One can see clearly the location and height of the bump and also how it arises from an increase in the success probability as the input amplitude $|\alpha|$ nears and exceeds $|\tilde\alpha|$, even as the fidelity begins to decrease.  Figure~\ref{fig6} plots the fidelity for the perfect amplifier on an expanded scale so one can see more clearly how it decreases as $|\alpha|$ nears and exceeds $|\tilde\alpha|$.

\subsection{Discussion of enhanced $N=1$ PFP}

The results in the previous subsection show clearly that the nondeterministic $\mu^2<1$-amplifiers can violate the na\"ive bound~(\ref{eq:PFPapproxbound}) and thus perform better, according to the PFP metric, than a deterministic ideal linear amplifier.  There are a couple of ways to think about this, both related to the status and interpretation of the approximate bound and, particularly, to the meaning of fidelity-based measures of amplifier performance, especially for small input amplitudes.

The first and most convincing response is that the ${\rm PFP}=1$ result for an ideal linear amplifier is not a strict bound even for deterministic devices.  As we have already seen in our discussion of the exact bound on $\mu^2$-amplifiers, one can do better on the PFP metric, at least for small $|\alpha|$, by targeting amplification with gain $G$, but actually doing ideal linear amplification with a smaller gain $g_2=G/g_1<G$.  The spreading of the output Gaussian is reduced by having smaller gain; in the fidelity, this reduction more than compensates, for small $|\alpha|$, for the fidelity reduction that comes from not centering the output Gaussian on the target amplitude.  This effect on fidelity by not amplifying up to the target state lies behind the fidelity enhancements for linear amplification and continuous variable teleportation reported in the literature~\cite{Namiki2011a,Chiribella2013a,CochRalp03}.

The first lesson here is that the $\mbox{PFP}=1$ benchmark for amplifier performance should not be treated as an absolute bound, especially for small input amplitudes; as we have already seen, $\mu^2$-amplifiers can beat the approximate bound, but they cannot beat the PFP$_0$ bound that is achieved by replacing the immaculate stage of the $\mu^2$-amplifier with doing nothing.  A second lesson has more far-reaching consequences: since the fidelity enhancement that comes from not amplifying up to the target state has nothing to do with the traditional amplifier goal of preserving signal in the face of noise, we are prompted to view fidelity-based measures with suspicion, as misleading indicators for evaluating amplifier performance, especially for small input amplitudes~\cite{ComCav13}.

\begin{figure}[hbtp]
\includegraphics[width=0.45\textwidth]{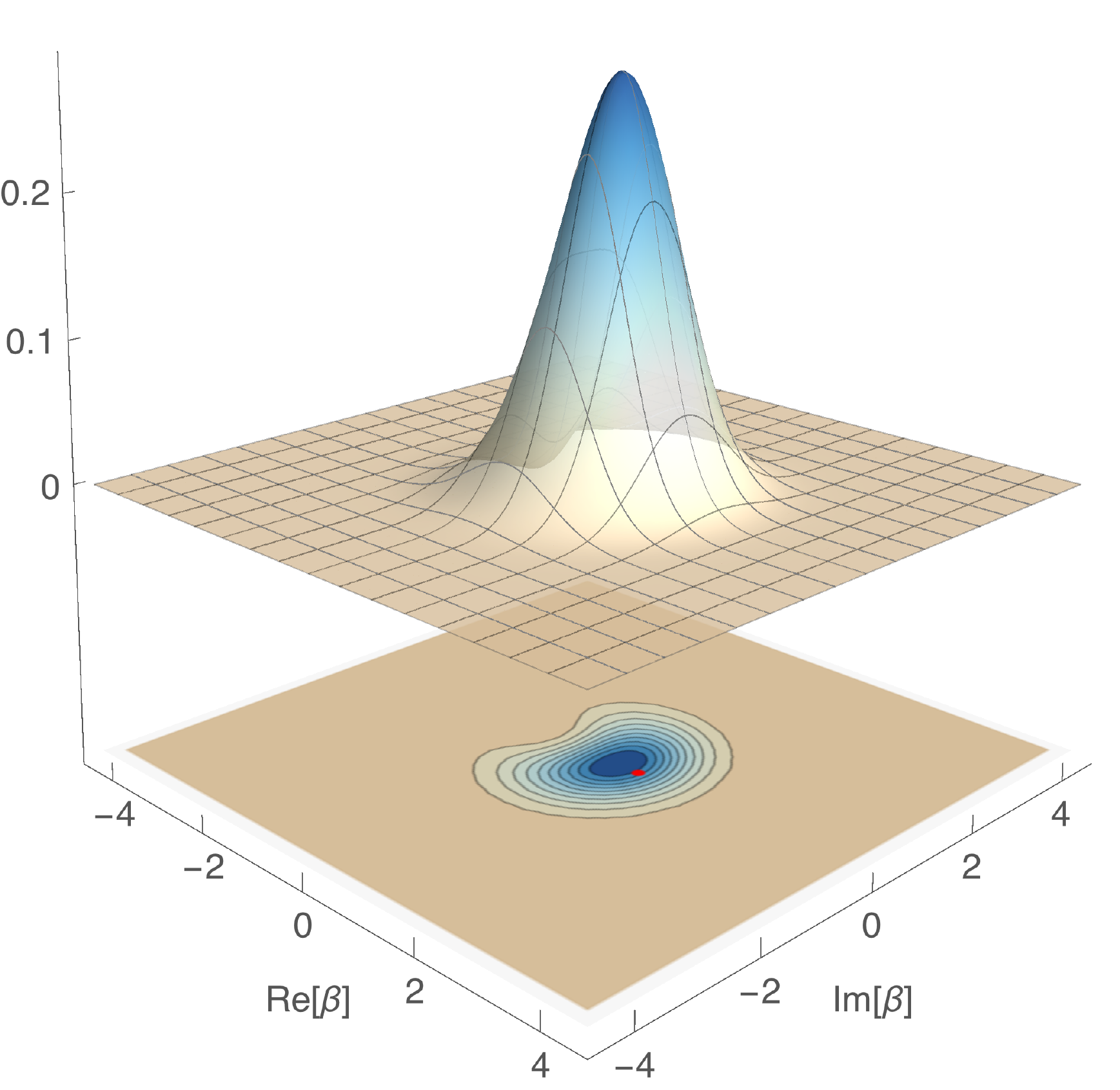}
\caption{(Color online) $Q$-function, $Q(\beta)$, of the output of an immaculate amplifier ($\mu^2=0$) with $N=1$, $G=9$, and input coherent-state amplitude $\alpha=|\alpha_{\mathrm{bump}}|=0.988|\tilde\alpha|= 0.110$. A red dot marks the mean amplitude of the target coherent state, demonstrating the extent to which the mean amplitude of the actual output differs.}\label{fig7}
\end{figure}

\begin{figure}[htbp]
\includegraphics[width=0.45\textwidth]{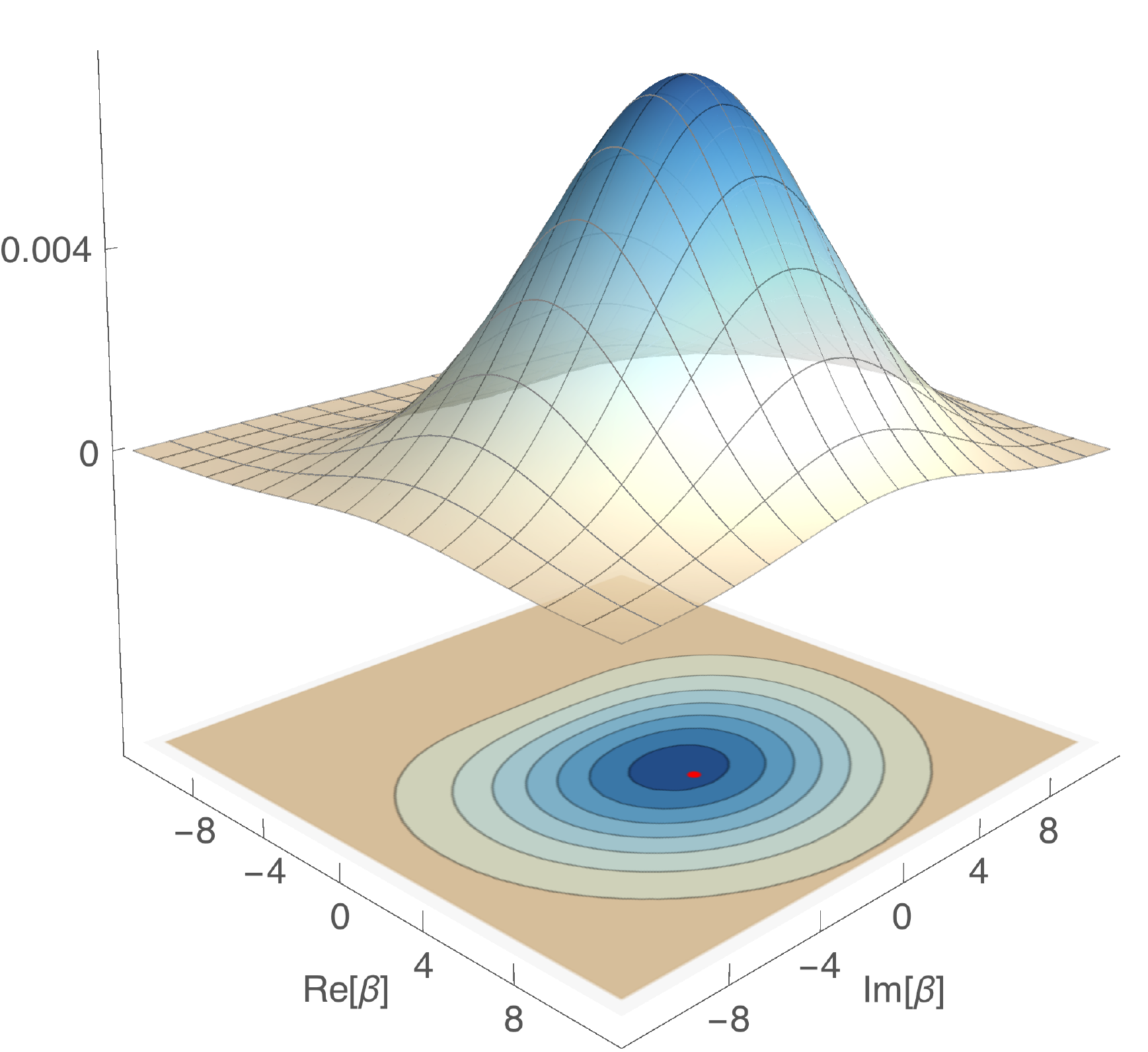}
\caption{(Color online) $Q$-function, $Q(\beta)$, of the output of a perfect amplifier ($\mu^2=1/2$) with $N=1$, $G=9$, and input coherent-state amplitude $\alpha=|\alpha_{\mathrm{bump}}|=0.572|\tilde\alpha|= 0.407$. A red dot marks the mean amplitude of the target coherent state.}\label{fig8}
\end{figure}

The second response comes from the nonGaussian character of the output states of our physical $\mu^2$-amplifiers.  The approximate bound~(\ref{eq:PFPapproxbound}) comes from an uncertainty-principle argument, more precisely, from the requirement~(\ref{eq:SNR_bound}) that the signal-to-noise resolvability not increase under probabilistic amplification.  The approximate bound ends up being expressed in terms of a probability-fidelity product because for Gaussian output states with symmetric noise, the fidelity to a coherent state centered on the Gaussian state is given by the same combination of parameters as the quadrature variances.  As we just discussed, however, the bump in PFP appears to be related to enhancing fidelity by not having the output state centered on the target coherent state; moreover, the output states of $\mu^2$-amplifiers that have enhanced PFP are nothing like Gaussian states with symmetric noise.  Both these considerations suggest that the connection of the SNR argument to the PFP is tenuous for states that have enhanced \hbox{PFP}.  They suggest again that we should view fidelity-based measures with suspicion and instead look at measures such as SNR that characterize amplifier performance directly.  Before turning to an examination of  the SNRs achieved by $\mu^2$-amplifiers, to see if these amplifiers provide any advantage over deterministic amplification, we consider briefly the output states of the $N=1$ immaculate and perfect amplifiers to highlight the properties just discussed.}

For immaculate amplification generically, the fidelity as a function of the
input amplitude decreases, while the success probability increases as a
function of input amplitude~\cite{ComCav13}.  The same is true for all
$\mu^2$-amplifiers, for $\mu^2<1$, as they are built around a first-stage
immaculate amplifier.  The plots in Fig.~\ref{fig5} illustrate this behavior
for $N=1$ immaculate and perfect amplifiers.  The bump region of input amplitudes
where the PFP exceeds unity is due to the fact that the success probability
starts to rise as $|\alpha|$ nears and exceeds $|\tilde\alpha|$; the fidelity
falls at the same place---and eventually falls precipitously---but not fast enough
at the beginning of the rise of $\Pcheck$ to prevent the PFP from exceeding unity.
The fall-off in fidelity is intuitive, since when $|\alpha|$ nears and exceeds
$|\tilde\alpha|=1/g_1$, the immaculate stage of the amplifier is at or beyond the
limit of the region where it can be said to be doing anything like amplification
to a target coherent state.  If one replots the perfect-amplifier fidelity by
itself for clarity (see Fig.~\ref{fig6}), one finds the unsurprising result
that the perfect amplifier's fidelity is, for small $|\alpha|$, larger than that
of an ideal amplifier with the same gain; the more important point is that the
roll-off of the fidelity is slower than for the comparable immaculate amplifier.

In the region of the PFP bump, the plots in Figs.~\ref{fig5} and~\ref{fig6} indicate that the fidelity to the target coherent state $\ket{G\alpha}$ is decreasing away from unity.  Fidelity is, however, a very poor indicator of what is happening to the output state in phase space; we can get a much better idea of what is happening by looking at the output state's $Q$-function.  Reference~\cite{ComCav13} investigated the output $Q$-function for immaculate amplifiers and found highly nonGaussian features for $|\alpha|\agt|\tilde\alpha|$.  Figures~\ref{fig7} and~\ref{fig8} plot the $Q$ functions of the output state of $N=1$ immaculate and perfect amplifiers when the input amplitude is chosen to be at the peak of the PFP bump, i.e., at $|\alpha|=|\alpha_{\rm bump}|$.  The outputs of both the immaculate and perfect amplifiers are distorted away from symmetric noise and display nonGaussian features; the disortion and nonGaussian features are less pronounced in the perfect amplifier because the noise added by the second state of ideal amplification tends to wash out the distortion and nonGaussian behavior.

\subsection{Exact SNR for $N=1$}

As the $G^2pF$ product has only a limited operational significance (the limitations
are inherited from fidelity itself), we turn to signal-to-noise ratio (SNR) as a figure of
merit. The SNR figure of merit has direct applications to quantum metrology.  Our goal in this
subsection is to see if the PFP bump corresponds to a similar advantage in signal-to-noise.

\begin{figure}[htbp]
\includegraphics[width=0.5\textwidth]{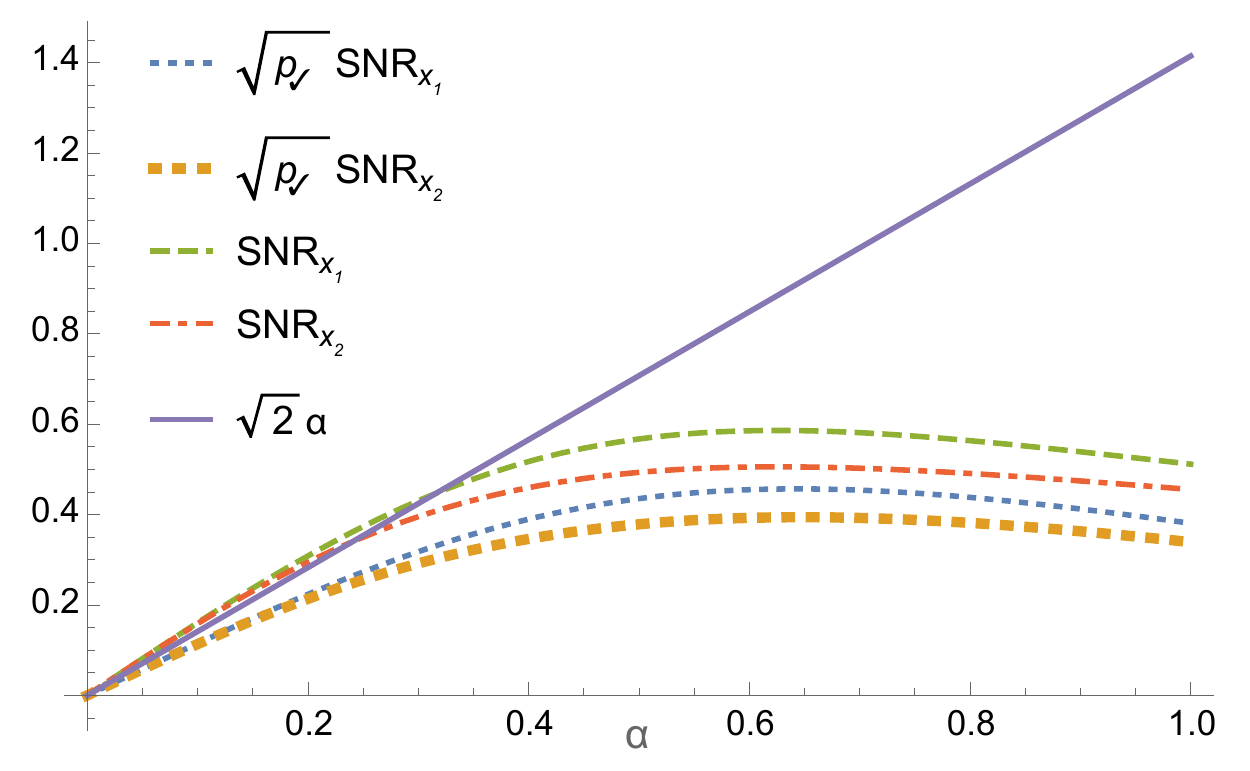}
\caption{(Color online) SNRs and root-probability--SNR products as a function of real input amplitude $\alpha$ for the amplitude ($x_1$) and phase ($x_2$) quadratures of the output state of a perfect amplifier ($\mu^2=1/2$) with $N=1$ and $G=9$.  Also plotted is the input SNR, which is given by $\sqrt{2}|\alpha|$.  The SNRs for the output state exceed the input SNR for small $\alpha$, but the root-probability--SNRs for the output do not exceed the input SNR.}\label{fig9}
\end{figure}

The SNR for a quadrature $q$  is defined as $\mathrm{SNR}_q =
\left<q\right>/\sqrt{V_q}$.  Henceforth, we assume moments are calculated
using antinormal ordering, which is the appropriate ordering if we imagine
measuring both quadratures.  In Fig.~\ref{fig9} we plot, for a perfect amplifier,
the SNR for both quadratures along with the $\sqrt{p_\checkmark}$SNR products and the SNR of the input
state given by $\sqrt{2}|\alpha|$; we see that although the raw SNR
surpasses that of the input state, once the success probability is included,
this is never the case. We also confirm the amplitude squeezing
observed in the $Q$-function (see Figs.~\ref{fig7} and \ref{fig8})
by  noting that the amplitude quadrature SNR is always greater than the phase quadrature.

\begin{figure}[htbp]
\includegraphics[width=0.45\textwidth]{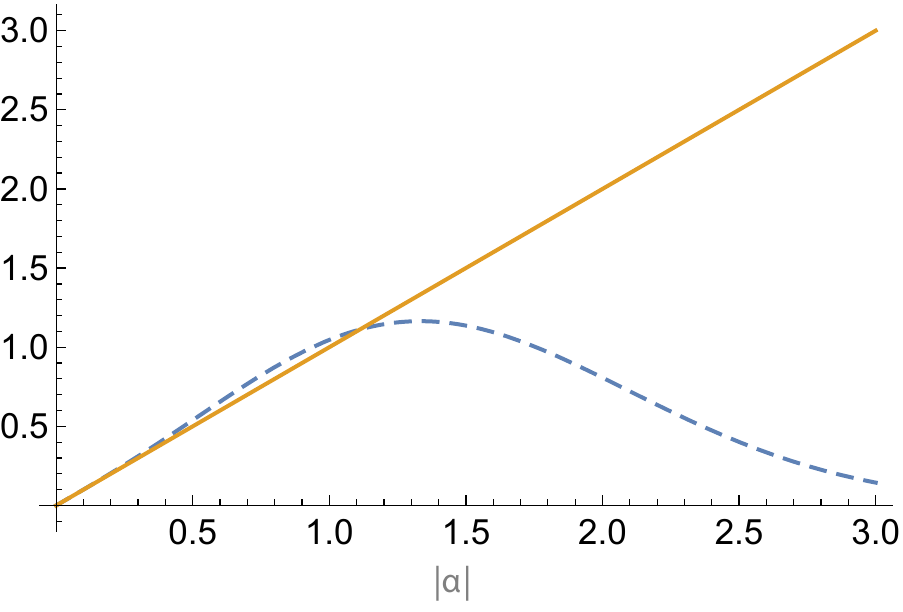}
\caption{(Color online) Number-based root-probability--SNR product as a function of input amplitude $|\alpha|$ for the output state of a perfect amplifier ($\mu^2=1/2$) with $N=1$ and $G=9$.  Also plotted is the input number-based SNR, which is given by $|\alpha|$.  The small enhancement of the root-probability--SNR is due to the squeezing of the amplitude quadrature.
}\label{fig10}
\end{figure}

Finally, one can also calculate a number-based SNR for the perfect amplifier.
In this case (see Fig.~\ref{fig10}) the $\sqrt{p_\checkmark}$SNR product does demonstrate an
improvement over the input.  A similar effect was found in~\cite{ComCav13}, and
here it should  be noted that the improvement is happening around the peak in
the success probability where the fidelity is declining.  As pointed out in
\cite{ComCav13}, this is a consequence of amplitude squeezing; i.e., it is
an effect of the amplifier not acting like a phase-preserving
linear amplifier.

\section{Conclusion}\label{sec:con}

In this paper we provide a physical construction of a family of reduced-noise, nondeterministic linear
amplifiers, which we call $\mu^2$-amplifiers.  The noise characteristics of these devices, as 
measured by $s$-ordered second moments of added noise, interpolate between an immaculate 
amplifier and an ideal amplifier as $\mu^2$ varies from 0 to 1.

Using an optimal physical realization of an immaculate amplifier, whose inputs
are restricted by a cutoff $N$ in the number basis, we bound
the performance of our proposed devices for all $N>1$.  For those who favor fidelity
to a target coherent state as a measure of performance, we bound performance in
terms of a probability-fidelity product, $G^2\Pcheck F$, and we show that
$G^2\Pcheck F\le1$ as long as $N\ge2$.  Similarly, we bound performance
in terms of the $s$-ordered noise figure, i.e., the ratio of input SNR to output root-probability--SNR.
Both types of bounds are saturated by an ideal linear amplifier.  To supplement these results,
we perform an exact analysis for $N=1$, where our bound on probability-fidelity product
can be violated.  Our exact results show that this violation is essentially spurious, raising
questions about fidelity as a performance metric for linear amplifiers instead of indicating
any particular utility for devices with $\mu^2<1$.  These
conclusions are strengthened by showing that the antinormally ordered root-probability--SNRs for
perfect amplifiers are not as good as for an ideal linear amplifier.

We leave a number of questions open for others to consider.  Although it is known that our construction is
optimal with respect to fidelity and working probability for an immaculate amplifier ($\mu^2=0$) and
with respect to added noise for an ideal linear amplifier ($\mu^2=1$), we have not proven
optimality with respect to any figure of merit for the intermediate values of $\mu^2$.
Another topic of interest would be to generalize our
results to phase-sensitive linear amplifiers~\cite{Caves1982a}.  Recently Namiki~\cite{Nam15}
has considered this with a Gaussian average, but it would be interesting to redo these calculations for reduced-noise amplifiers,
without having to resort to the Gaussian averaging.  Finally, alternative bounds, such as state-discrimination-based bounds~\cite{ComCav13,DunAnd12,ChirYangHuan15}, and alternative constructions, such as measurement-based realizations~\cite{CavComJiaPan12}, provide interesting avenues for further research.

While our analysis suggests that $\mu^2$-amplifiers have little utility for the traditional tasks of linear amplifiers,
they still hold promise for a range of other tasks, including those where immaculate amplification has already found application, such as quantum key distribution~\cite{Fiurasek:2012p7670,Walk:2013p403,Blandino:2012p5681} and the distillation of quantum correlations~\cite{ChrWalAss14}.  Other possibilities include offline preparation of resources for teleportation~\cite{Ralp11,Blandino:2016jd} and for processing of quantum information, where one might be willing to tolerate reasonably low success probabilities for the payoff of high-quality resource states.  Such applications are characterized by other performance metrics, such as key rate, than the fidelity and SNR-based measures considered in this paper.  Indeed, rare but high-quality resource states might be optimal under alternative cost metrics~\cite{CombFerr15}.  Finally, we note that although most analysis of nondeterministic amplifiers has focused on quantum applications and limits, there are applications at the other end of the spectrum in classical optics.  The use of immaculate-amplification ideas seems to be gaining traction in signal-processing applications~\cite{AtaiEsmaKuo15,Vasilyev15}.

\begin{acknowledgments}
The authors thank R\'emi Blandino, Zhang Jiang, and Shashank Pandey for helpful conversations.
This work was supported in part by US Office of Naval Research Grant No.~N00014-15-1-2167, US National Science Foundation Grant No.~PHY-1521106; ARC Centre of Excellence for Engineered Quantum Systems, CE110001013; and the ARC Centre of Excellence for Quantum Computation and Communication Technology, CE110001027. NW acknowledges support from the EPSRC (through the NQIT Quantum Hub).
\end{acknowledgments}

\bibliography{mu_amp}

\end{document}